\documentclass[
    ,final            
  ]
  {aipproc}

\layoutstyle{6x9}

\def\hermes{{\sc Hermes}}

\def\ubar{\ensuremath{\overline{u}}}
\def\dbar{\ensuremath{\overline{d}}}

\begin{document}

\title{Measurement of $\Delta S$ in the nucleon at HERMES from semi-inclusive
DIS}


\author{H. E. Jackson}{
  address={Argonne National Laboratory},
address={on behalf of
the HERMES collaboration}}

\begin{abstract}
The helicity density of the strange quark sea in the proton has been
extracted from measurements of polarized semi-inclusive production of
charged kaons in deep-inelastic scattering of positrons
from a polarized deuteron
target. In the region of measurement of x $>$ 0.02 the helicity density
is zero within experimental error and the measured first moment of the 
density is 0.006$\pm$0.029(stat.)$\pm$0.007(sys.). The first moment 
of the axial charge
in the measured region is substantially less than that inferred from 
hyperon semi-leptonic decays.
\end{abstract}

\maketitle


\section{Introduction}

The helicity distribution of the strange quark sea is of great interest as 
a probe of the spin properties of the quark sea in the nucleon. Because 
strange quarks carry no isospin, the total strange quark helicity density
$\Delta S(x)\equiv\Delta s(x)+\Delta\overline{s}(x)$ can be extracted from
measurements of scattering off the deuteron alone which is isoscalar. In
effect, measurements of the inclusive spin asymmetries provide an estimate
of the helicity density $\Delta Q(x)\equiv\Delta u(x)+\Delta\ubar (x)+
\Delta d(x)+\Delta\dbar (x)$ of the non-strange sea. Using the spin asymmetries
measured for the charged kaons as the second measured quantity, it is
possible to extract $\Delta S(x)$.  By measuring the charged kaon
multiplicites with the same data set, the fragmentation functions relevant
to the extraction process can be obtained without resort to other experiments.
Aside from that of isospin symmetry  between the proton and the neutron,
the only assumption required in the analysis is charge-conjugation
invariance in the fragmentation process. A precise ``isoscalar''
extraction of $\Delta S(x)$ using semi-inclusive DIS on the deuteron at 
{\hermes}~\cite{hermes:deltaq} is presented 
here.

\section{The experiment}

In leading-order (LO), the semi-inclusive
virtual photon double-spin asymmetry $A_1^K$ in semi-inclusive
production of a Kaon is given by
\begin{equation}
      A_1^K(x,z)={\sigma_{1/2}^K -\sigma_{3/2}^K\over\sigma_{1/2}^K
+\sigma_{3/2}^K}=
{\sum_q e_q^2\ \Delta q(x)\ D_q^K(z)\over
      \sum_q e_q^2\ q(x)\ D_q^K(z)}\,,
\label{eq:hadronasym}
\end{equation}
The asymmetries for the analysis reported here were recorded by the
\hermes\ experiment using a longitudinally nuclear-polarized deuteron
gas target internal to the E = 27.6 GeV {\sc Hera} positron storage ring at
{\sc Desy}. The self-induced beam polarization is
measured continuously with Compton back-scattering of circularly polarized
laser beams~\cite{hermes:tpol2,hermes:lpol}.
The open-ended target cell is fed by an
atomic-beam source based on Stern-Gerlach separation~\cite{hermes:ABS} with
hyperfine transitions.
The nuclear polarization of the atoms is flipped at 90\,s time intervals,
while both this polarization and the atomic fraction inside the target cell are
continuously measured~\cite{hermes:BRP,hermes:TGA}.

Scattered beam leptons and coincident hadrons are detected by
the \hermes\ spectrometer~\cite{hermes:spectr}. Leptons are identified
with an efficiency exceeding 98\% and a hadron
contamination of less than 1\% using
an electromagnetic calorimeter, a transition-radiation detector, a
preshower scintillation counter and a {\v C}erenkov detector.
Charged kaons are identified using a dual-radiator ring-imaging
{\v C}erenkov detector~\cite{hermes:rich}.
Events were selected subject to the kinematic
requirements $Q^2>1$\,GeV$^2$, $W^2 > 10$\,GeV$^2$ and $y < 0.85$,
where
$W$ is the invariant mass of the
photon-nucleon system, $\nu$ is the virtual photon energy, and $y=\nu/E$.
Coincident hadrons were accepted if $0.2<z<0.8$ where $z=E_{hadron}/\nu$,
and $x_F\approx 2p_L/W>0.1$, where $p_L$ is the longitudinal
momentum of the hadron with respect to the photon
direction in the photon-nucleon center of mass frame.

\section{Extraction of helicity distributions}

For the deuteron Eq.~\ref{eq:hadronasym} reduces to a simple form which
reflects its isoscalar character
\begin{eqnarray}\label{parton6}
A^K_1(x) & = & \frac{
\Delta Q(x)\int {\cal D}^K_{Q}(z)dz +\Delta S(x)\int {\cal D}_{S}^K(z)dz}{
Q(x)\int {\cal D}^K_{Q}(z)dz +S(x)\int {\cal D}_{S}^K(z)dz}
\end{eqnarray}
Where
$\int {\cal D}^K_{Q}(z)dz  =  4\int D(z)_u^{K^\pm}dz+
\int D(z)_d^{K\pm}dz $ and $
\int{\cal D}^K_{S}(z)dz = 2\int D(z)_s^{K^\pm}dz.$ 
The corresponding inclusive asymmetry is given by
\begin{equation}\label{parton7}
A_1(x) =  \frac{
5\Delta Q(x)+2\Delta S(x)}{5Q(x)+2S(x)}.
\end{equation}

\begin{figure}[h]
\includegraphics[width=6cm]{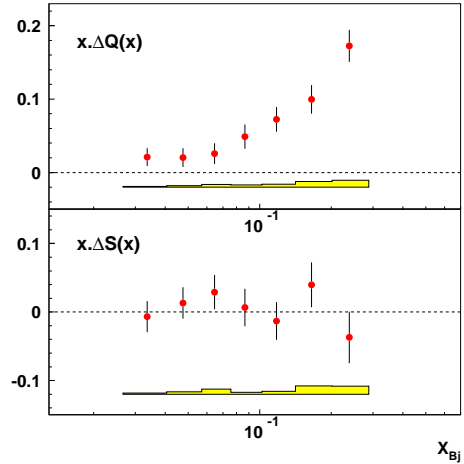}
\caption{\label{fig:helicities} Strange and non-strange quark helicity
distributions at $\langle Q^2\rangle=2.5$\,GeV$^2$, as a function of
Bjorken $x$.
The error bars are statistical, and the bands at the
bottom represent the systematic uncertainties.}
\end{figure}

The helicity distributions $\Delta Q(x)$ and $\Delta S(x)$
have been extracted directly from the measured values of
$A_{1}(x)$ and $A^{K^{\pm}}_{1}(x)$ using the fragmentation functions 
defined above as measured with the same data set and the 
parton distributions Q(x) and S(x) taken from the latest
compilations.
The resulting strange and non-strange
helicity distributions weighted by Bjorken $x$ are presented in
Fig.~\ref{fig:helicities}.
The non-strange helicity distribution is in excellent agreement with
that derived from the published five-component flavor
decomposition~\cite{hermes:deltaq} of the proton helicities. While of
much improved precision and free of the systematic uncertainties in
the fragmentation functions, the strange helicity distribution also
agrees well with the results reported therein, and is consistent with
zero over the measured range.

The first moment of $\Delta S(x)$ in the measured region is 0.006$\pm$
0.029(stat.)$\pm$0.007(sys.).
The result for the first moment of $\Delta Q(x)$ is 0.286$\pm$0.026(stat.)
$\pm$ 0.011(sys.) which is in excellent agreement with values
previously measured at HERMES~\cite{hermes:dq1999,hermes:deltaq}. The first
moment of $\Delta S(x)$ is consistent with zero in the measured region.
Because the very small density of strange quarks above $x=0.3$, the
contribution of any non-zero helicity density in this region is negligible
compared to the systematic error of the measurement. Consequently, the value
for S(x) can be safely taken as the moment over the
Bjorken x range 0.02-1.0. The vanishing values recently
reported~\cite{compass:g1d} for $g_{1,d}(x)$
at lower values of x, suggest that any
contribution to the first moment of $\Delta Q(x)$
below $x=0.02$ will be very small. While an anomalously large contribution
to the strange moment at $x<0.02$ can not be ruled out, the data reported
here strongly suggest that the first moment vanishes. 
If true the violation of the Ellis-Jaffe sum rule observed
in inclusive scattering is not due to a significant negative polarization
of the strange sea. The result for the first moment of the octet axial 
charge $\Delta a_8=\int (\Delta Q(x)-2\Delta S(x))dx$ is 
0.274$\pm$0.026(stat.)$\pm$ 0.011(sys.), substantially less than the value
inferred from hyperon decay.


\begin{theacknowledgments}
We gratefully acknowledge the {\sc Desy} management for its support, the staff
at {\sc Desy} and the collaborating institutions for the significant effort,
and our national funding agencies for their financial support.

\end{theacknowledgments}



\bibliographystyle{aipproc}   

\bibliography{dsletter}

\begin{thebibliography}{25}
\expandafter\ifx\csname natexlab\endcsname\relax\def\natexlab#1{#1}\fi
\expandafter\ifx\csname bibnamefont\endcsname\relax
  \def\bibnamefont#1{#1}\fi
\expandafter\ifx\csname bibfnamefont\endcsname\relax
  \def\bibfnamefont#1{#1}\fi
\expandafter\ifx\csname citenamefont\endcsname\relax
  \def\citenamefont#1{#1}\fi
\expandafter\ifx\csname url\endcsname\relax
  \def\url#1{\texttt{#1}}\fi
\expandafter\ifx\csname urlprefix\endcsname\relax\def\urlprefix{URL }\fi
\providecommand{\bibinfo}[2]{#2}
\providecommand{\eprint}[2][]{\url{#2}}

\bibitem[{\citenamefont{Lai et~al.}(2000)}]{pdf:cteq5}
\bibinfo{author}{\bibfnamefont{H.~L.} \bibnamefont{Lai}} \bibnamefont{et~al.},
  \bibinfo{journal}{Eur.~Phys.~J.~C} \textbf{\bibinfo{volume}{12}},
  \bibinfo{pages}{375} (\bibinfo{year}{2000}).

\bibitem[{\citenamefont{Martin et~al.}(2002)}]{pdf:MRST2001}
\bibinfo{author}{\bibfnamefont{A.~D.} \bibnamefont{Martin}}
  \bibnamefont{et~al.}, \bibinfo{journal}{Eur. Phys. J.~C}
  \textbf{\bibinfo{volume}{23}}, \bibinfo{pages}{73} (\bibinfo{year}{2002}).

\bibitem[{\citenamefont{Ashman et~al.}(1988)}]{emc:g1}
\bibinfo{author}{\bibfnamefont{J.}~\bibnamefont{Ashman}} \bibnamefont{et~al.}
  (\bibinfo{collaboration}{EMC}), \bibinfo{journal}{Phys. Lett.}
  \textbf{\bibinfo{volume}{B206}}, \bibinfo{pages}{364} (\bibinfo{year}{1988}).

\bibitem[{\citenamefont{Adams et~al.}(1997)}]{smc:g1p-2}
\bibinfo{author}{\bibfnamefont{D.}~\bibnamefont{Adams}} \bibnamefont{et~al.}
  (\bibinfo{collaboration}{SMC}), \bibinfo{journal}{Phys. Rev. D}
  \textbf{\bibinfo{volume}{56}}, \bibinfo{pages}{5330} (\bibinfo{year}{1997}).

\bibitem[{\citenamefont{Airapetian et~al.}(2005)}]{hermes:deltaq}
\bibinfo{author}{\bibfnamefont{A.}~\bibnamefont{Airapetian}}
  \bibnamefont{et~al.} (\bibinfo{collaboration}{\hermes}),
  \bibinfo{journal}{Phys.~Rev.~D} \textbf{\bibinfo{volume}{71}},
  \bibinfo{pages}{012003} (\bibinfo{year}{2005}).

\bibitem[{\citenamefont{Barber et~al.}(1994)}]{hermes:tpol2}
\bibinfo{author}{\bibfnamefont{D.~P.} \bibnamefont{Barber}}
  \bibnamefont{et~al.}, \bibinfo{journal}{Nucl. Inst. \& Meth.}
  \textbf{\bibinfo{volume}{A 338}}, \bibinfo{pages}{166}
  (\bibinfo{year}{1994}).

\bibitem[{\citenamefont{Beckmann et~al.}(2002)}]{hermes:lpol}
\bibinfo{author}{\bibfnamefont{M.}~\bibnamefont{Beckmann}}
  \bibnamefont{et~al.}, \bibinfo{journal}{Nucl. Inst. \& Meth.}
  \textbf{\bibinfo{volume}{A 479}}, \bibinfo{pages}{334}
  (\bibinfo{year}{2002}).

\bibitem[{\citenamefont{Stock et~al.}(1994)}]{hermes:ABS}
\bibinfo{author}{\bibfnamefont{F.}~\bibnamefont{Stock}} \bibnamefont{et~al.},
  \bibinfo{journal}{Nucl. Inst. \& Meth.} \textbf{\bibinfo{volume}{A 343}},
  \bibinfo{pages}{334} (\bibinfo{year}{1994}).

\bibitem[{\citenamefont{Baumgarten et~al.}(2002)}]{hermes:BRP}
\bibinfo{author}{\bibfnamefont{C.}~\bibnamefont{Baumgarten}}
  \bibnamefont{et~al.}, \bibinfo{journal}{Nucl. Inst. \& Meth.}
  \textbf{\bibinfo{volume}{A 482}}, \bibinfo{pages}{606}
  (\bibinfo{year}{2002}).

\bibitem[{\citenamefont{Baumgarten et~al.}(2003)}]{hermes:TGA}
\bibinfo{author}{\bibfnamefont{C.}~\bibnamefont{Baumgarten}}
  \bibnamefont{et~al.}, \bibinfo{journal}{Nucl. Inst. \& Meth.}
  \textbf{\bibinfo{volume}{A 496}}, \bibinfo{pages}{263}
  (\bibinfo{year}{2003}).

\bibitem[{\citenamefont{Ackerstaff et~al.}(1998)}]{hermes:spectr}
\bibinfo{author}{\bibfnamefont{K.}~\bibnamefont{Ackerstaff}}
  \bibnamefont{et~al.} (\bibinfo{collaboration}{\hermes}),
  \bibinfo{journal}{Nucl. Inst. \& Meth.} \textbf{\bibinfo{volume}{A 417}},
  \bibinfo{pages}{230} (\bibinfo{year}{1998}).

\bibitem[{\citenamefont{Akopov et~al.}(2002)}]{hermes:rich}
\bibinfo{author}{\bibfnamefont{N.}~\bibnamefont{Akopov}} \bibnamefont{et~al.},
  \bibinfo{journal}{Nucl. Inst. \& Meth.} \textbf{\bibinfo{volume}{A 479}},
  \bibinfo{pages}{511} (\bibinfo{year}{2002}).

\bibitem[{\citenamefont{Airapetian et~al.}(1998)}]{hermes:g1p}
\bibinfo{author}{\bibfnamefont{A.}~\bibnamefont{Airapetian}}
  \bibnamefont{et~al.} (\bibinfo{collaboration}{\hermes}),
  \bibinfo{journal}{Phys. Lett.} \textbf{\bibinfo{volume}{B442}},
  \bibinfo{pages}{484} (\bibinfo{year}{1998}).

\bibitem[{\citenamefont{Mankiewicz et~al.}(1992)}]{ph:pepsi}
\bibinfo{author}{\bibfnamefont{L.}~\bibnamefont{Mankiewicz}}
  \bibnamefont{et~al.}, \bibinfo{journal}{Comp. Phys. Comm.}
  \textbf{\bibinfo{volume}{71}}, \bibinfo{pages}{305} (\bibinfo{year}{1992}).

\bibitem[{\citenamefont{Akushevich et~al.}(1998)}]{hermes:radgen}
\bibinfo{author}{\bibfnamefont{I.}~\bibnamefont{Akushevich}}
  \bibnamefont{et~al.} (\bibinfo{year}{1998}),
  \eprint[http://arXiv.org/abs]{hep-ph/9906408}.

\bibitem[{\citenamefont{Sj{\"o}strand et~al.}(2001)}]{ph:jetset}
\bibinfo{author}{\bibfnamefont{T.}~\bibnamefont{Sj{\"o}strand}}
  \bibnamefont{et~al.}, \bibinfo{journal}{Comp.~Phys.~Comm.}
  \textbf{\bibinfo{volume}{135}}, \bibinfo{pages}{238} (\bibinfo{year}{2001}).

\bibitem[{\citenamefont{Anthony et~al.}(2003)}]{E155:g2}
\bibinfo{author}{\bibfnamefont{P.~L.} \bibnamefont{Anthony}}
  \bibnamefont{et~al.} (\bibinfo{collaboration}{E155}), \bibinfo{journal}{Phys.
  Lett.} \textbf{\bibinfo{volume}{B553}}, \bibinfo{pages}{18}
  (\bibinfo{year}{2003}).

\bibitem[{\citenamefont{Kretzer}(2000)}]{frag:kretzer}
\bibinfo{author}{\bibfnamefont{S.}~\bibnamefont{Kretzer}},
  \bibinfo{journal}{Phys. Rev. D} \textbf{\bibinfo{volume}{62}},
  \bibinfo{pages}{054001} (\bibinfo{year}{2000}).

\bibitem[{\citenamefont{Kniehl et~al.}(2000)}]{frag:kkp}
\bibinfo{author}{\bibfnamefont{B.~A.} \bibnamefont{Kniehl}}
  \bibnamefont{et~al.}, \bibinfo{journal}{Nucl.~Phys.}
  \textbf{\bibinfo{volume}{B582}}, \bibinfo{pages}{514} (\bibinfo{year}{2000}).

\bibitem[{\citenamefont{Pumplin et~al.}(2002)}]{pdf:cteq6}
\bibinfo{author}{\bibfnamefont{J.}~\bibnamefont{Pumplin}} \bibnamefont{et~al.},
  \bibinfo{journal}{J.~High~Energy~Phys.} \textbf{\bibinfo{volume}{7}},
  \bibinfo{pages}{12} (\bibinfo{year}{2002}).

\bibitem[{\citenamefont{Ackerstaff et~al.}(1999)}]{hermes:dq1999}
\bibinfo{author}{\bibfnamefont{K.}~\bibnamefont{Ackerstaff}}
  \bibnamefont{et~al.} (\bibinfo{collaboration}{\hermes}),
  \bibinfo{journal}{Phys. Lett.} \textbf{\bibinfo{volume}{B464}},
  \bibinfo{pages}{123} (\bibinfo{year}{1999}).

\bibitem[{\citenamefont{Ageev et~al.}(2005)}]{compass:g1d}
\bibinfo{author}{\bibfnamefont{E.~S.} \bibnamefont{Ageev}} \bibnamefont{et~al.}
  (\bibinfo{collaboration}{\compass}), \bibinfo{journal}{Phys. Lett.}
  \textbf{\bibinfo{volume}{B612}}, \bibinfo{pages}{154} (\bibinfo{year}{2005}).

\bibitem[{\citenamefont{Lichtenstadt and Lipkin}(1995)}]{ph:lichten-lipkin}
\bibinfo{author}{\bibfnamefont{J.}~\bibnamefont{Lichtenstadt}}
  \bibnamefont{and} \bibinfo{author}{\bibfnamefont{H.~J.}
  \bibnamefont{Lipkin}}, \bibinfo{journal}{Phys. Lett.}
  \textbf{\bibinfo{volume}{B353}}, \bibinfo{pages}{119} (\bibinfo{year}{1995}).

\bibitem[{\citenamefont{Leader et~al.}(2000)}]{ph:leader}
\bibinfo{author}{\bibfnamefont{E.}~\bibnamefont{Leader}} \bibnamefont{et~al.},
  \bibinfo{journal}{Phys. Lett.} \textbf{\bibinfo{volume}{B488}},
  \bibinfo{pages}{283} (\bibinfo{year}{2000}).

\bibitem[{\citenamefont{Adeva et~al.}(1998)}]{smc:g1p-g1d}
\bibinfo{author}{\bibfnamefont{B.}~\bibnamefont{Adeva}} \bibnamefont{et~al.}
  (\bibinfo{collaboration}{SMC}), \bibinfo{journal}{Phys. Rev. D}
  \textbf{\bibinfo{volume}{58}}, \bibinfo{pages}{112001}
  (\bibinfo{year}{1998}).

\end{thebibliography}

\IfFileExists{\jobname.bbl}{}
 {\typeout{}
  \typeout{******************************************}
  \typeout{** Please run "bibtex \jobname" to optain}
  \typeout{** the bibliography and then re-run LaTeX}
  \typeout{** twice to fix the references!}
  \typeout{******************************************}
  \typeout{}
 }

\end{document}